# Electrical controlled rheology of a suspension of weakly conducting particles in dielectric liquid

Q. Guegan[1, a], J.-N. Foulc[1, b], F. Ayela[2, c] and O. Tillement[3, d]

[1]Laboratoire d'Electrostatique et de Matériaux Diélectriques (LEMD) / CNRS
25 av. des Martyrs - BP 166 - 38042 Grenoble Cedex 9, France

[2]Centre de Recherches sur les Très Basses Températures (CRTBT) / CNRS
25 av. des Martyrs - BP 166 - 38042 Grenoble Cedex 9, France

[3] Laboratoire de Physico-Chimie des Matériaux Luminescents (LPCML) / CNRS
Université Claude Bernard Lyon 1 - Domaine Scientifique de la Doua
10 rue A. M. Ampère - 69622 Villeurbanne Cedex, France

[a]quentin.guegan@grenoble.cnrs.fr, [b]jean-numa.foulc@grenoble.cnrs.fr,
[c]frederic.ayela@grenoble.cnrs.fr, [d]tillement@pcml.univ-lyon1.fr

**Keywords:** electrorheology, nanosupsension, dielectric materials, electric field, van der Waals, electrostatic forces.

**Abstract.** The properties of suspensions of fine particles in dielectric liquid (electrorheological fluids) subjected to an electric field lead to a drastic change of the apparent viscosity of the fluid. For high applied fields (~ 3-5 kV/mm) the suspension congeals to a solid gel (particles fibrillate span the electrode gap) having a finite yield stress. For moderate fields the viscosity of the suspension is continuously controlled by the electric field strength. We have proposed that in DC voltage the field distribution in the solid (particles) and liquid phases of the suspension and so the attractive induced forces between particles and the yield stress of the suspension are controlled by the conductivities of the both materials. In this paper we report investigation and results obtained with nanoelectrorheological suspensions: synthesis of coated nanoparticles (size ~ 50 to 600 nm, materials $Gd_2O_3$:Tb, $SiO_x$...), preparation of ER fluids (nanoparticles mixed in silicone oil), electrical and rheological characterization of the ER fluids. We also propose a possible explanation of the enhanced ER effect (giant ER fluids) taking into account the combined effects of the (nano)size of the particles, the Van der Waals forces between particles in contact and the electrostatic pressure in a very thin layer of insulating liquid.

**Introduction**

**Electrorheological Fluids.** Electrorheological (ER) fluids are concentrated suspensions of small solid particles in a dielectric liquid. These fluids are all distinguished by the ability to control or to stop their flow when an electric field (~1-5 kV/mm) is applied. This change of the rheological properties of the fluids is due to the fibres formation in the direction of the electric field. Those fibres appear within a few milliseconds. A few minutes later, the fibres gather to form thick columns. Moreover, the effect is totally reversible: when the field is turn off, the ER fluids return to liquid state.

ER fluids have been discovered in 1947 by W. M. Winslow [1,2]. They caused a great interest as they offered numerous applications in various domains (dampers, clutches, valves, actuators, artificial members...) with many advantages (speed, low electric consumption, simplicity...).

**Characterization.** The characterization of an ER fluid is represented by a graphic showing the shear stress τ as a function of the shear rate $\dot{\gamma}$. Fig. 3 presents a schematic characterization of an ER fluid.

Under electric field, we can distinguish two states: a solid state and a liquid one. That kind of behaviour can be adequately describes as Bingham bodies that also have a threshold (yield stress) upon which the liquid flows.

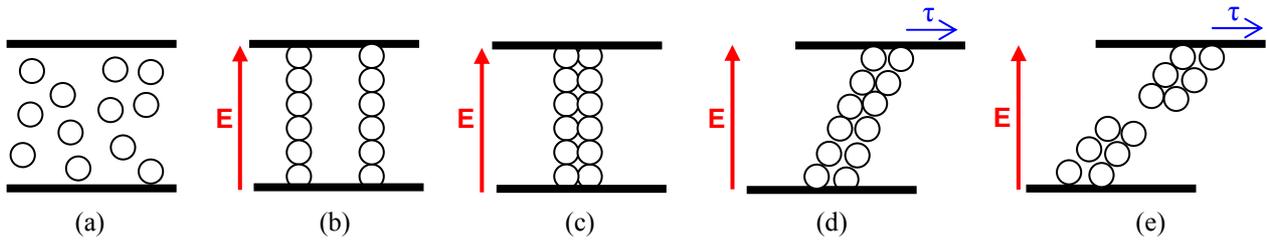

(a)  (b)  (c)  (d)  (e)

**Figure 1:** Illustration of the structure formation in ER fluids. E (applied electric field), τ (applied shear stress) and $τ_S$ (yield stress).
E = 0 and τ = 0 : particles are randomly distributed (a)
E ≠ 0 and τ = 0 : single-chaine structure (b), column structure (c)
E ≠ 0 and τ ≠ 0 : τ << $τ_S$ , the column is sloped (d) ; τ > $τ_S$ , the column is broken (e)

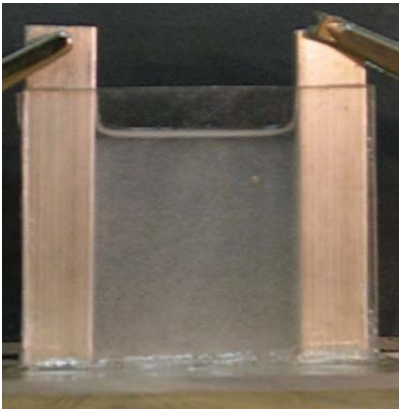
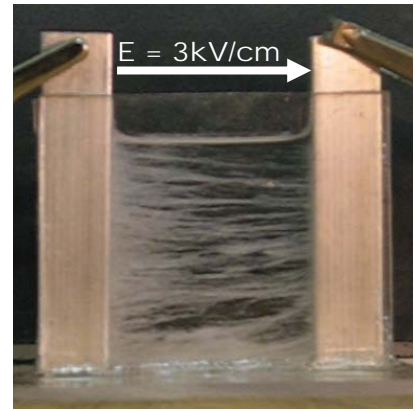

**Figure 2:** Suspension of cellulose dispersed in silicon oil. Without electric field (left) and, after a few seconds, with a 3 kV/cm field (right). We can obviously observe the formation of columns.

The strength (or the quality) of an ER fluid can be evaluated by the value of its static yield stress $τ_s$ which corresponds to the maximum force the ER fluid can handle before it breaks (Fig. 1(e)). Common ER fluids have a static yield stress about 5-8 kPa which is not sufficient for the majority of applications.

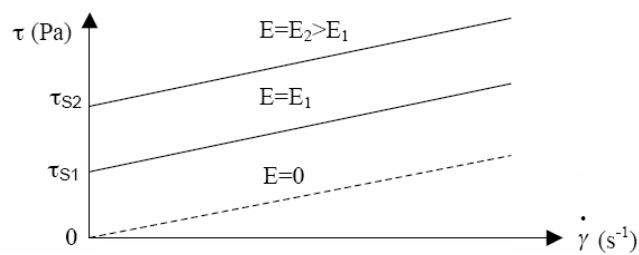

**Figure 3:** Illustration of the rheological behaviour of an ER fluid and the effect of the applied electric field E on the yield stress $τ_s$

**The giant ER effect.** Recently Wen *et al* announced [3] the synthesis of a new ER fluid which reach a yield stress of 130 kPa (20 times higher than traditional ER fluids) and all applications quoted above appears now to be realizable. This modified the situation to a significant extent of the interest of these fluids.

The main difference between their giant ER fluid and the common fluids is the size of the particles. Whereas classical fluids were used to be synthesize with micro-sized particles, Wen *et al* used nano-sized particles of BaTi03 (diameter 50 nm) coated with urea (10 nm thick).

**Tested fluids.** The yield stress of various fluids have been measured with a cylindrical rheometer (rheomat 115A) while the current density values were collected with a pico-ammeter (Fig. 4).

Fig. 5(a),(b) present measurements of ER fluids which consists of nano-sized $S_iO_x$ particles functionalized on the surface with 40% of amine groups and 60% of hydroxyl groups and dispersed in silicon oil (Wacker AS 4). Fig. 5(c) shows data from a dispersion of $S_iO_x$ particles in silicon oil. All fluids possess a volume fraction approximately equal to 15%. Measures were made at room temperature.

According to those data, the smaller the particles are, the higher the yield stress is (Fig. 5(a), (c)). In the same time, the current increases with the yield stress stress (Fig. 5(b)), which is in good agreement with the conduction model (see the next paragraph).

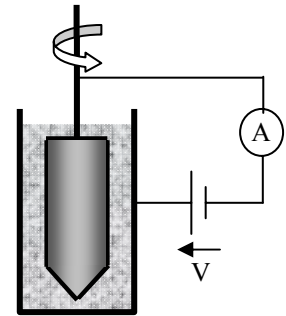

**Figure 4:** ER fluid measurement system

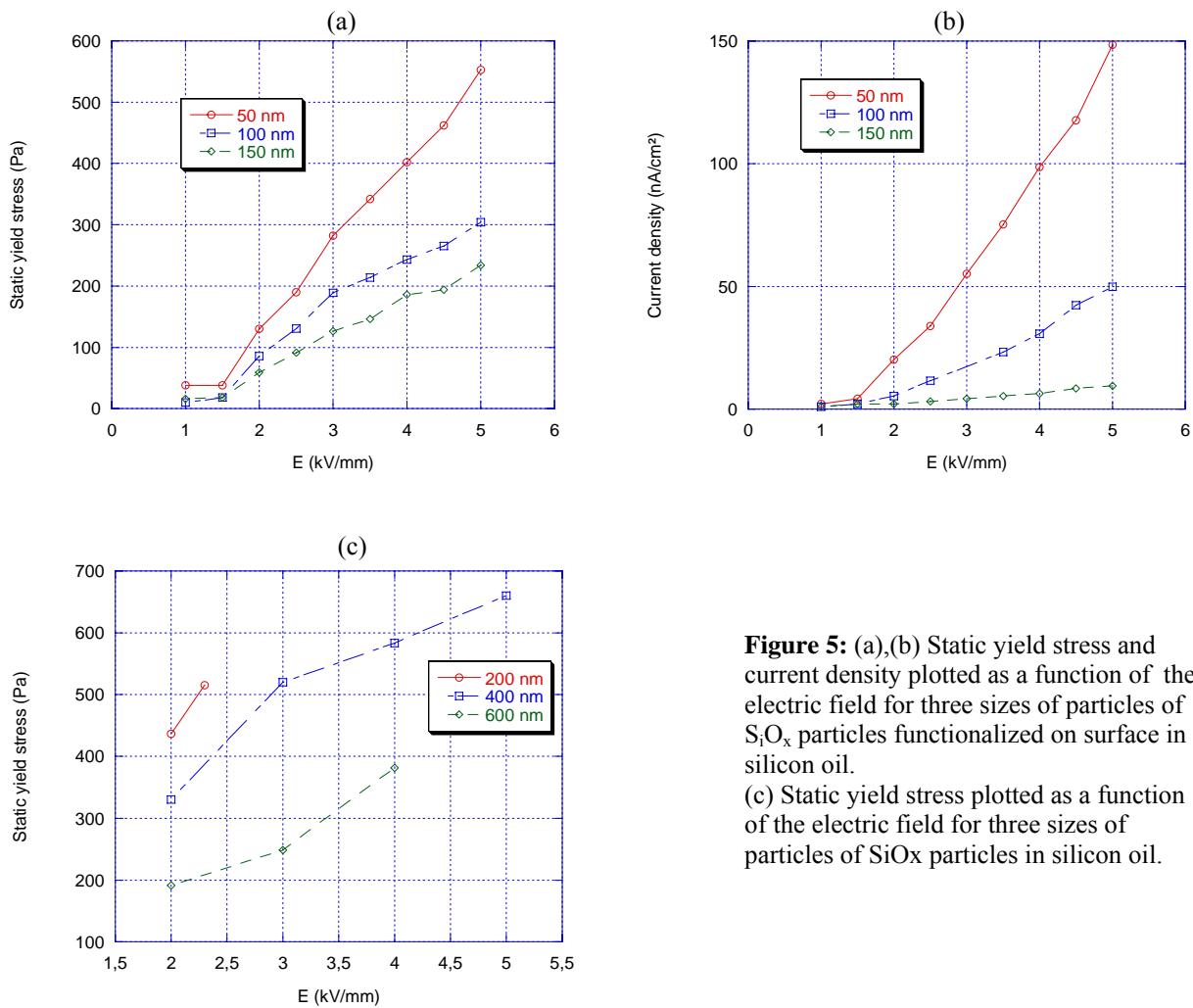

**Figure 5:** (a),(b) Static yield stress and current density plotted as a function of the electric field for three sizes of particles of $S_iO_x$ particles functionalized on surface in silicon oil.
(c) Static yield stress plotted as a function of the electric field for three sizes of particles of SiOx particles in silicon oil.

## Model for traditional ER fluids under DC electric field

**Two models.** Until 2003 and the discovery of the giant ER effect, two principal models were proposed to describe the ER phenomenon: the polarization and the conduction models. The polarization model introduces the permittivities of the different phases as the main parameters ($\varepsilon_{solid} > \varepsilon_{liquid}$). In the case of the conduction model, the difference of the conductivities between the solid and the liquid are used. Theory and experience give the following conditions for a fluid to be electrorheological: $(10^2$ to $10^3)\sigma_L \leq \sigma_S \leq 10^{-7}$-$10^{-5}$ S/m [4,5].

According to the data given by Wen *et al* [3], it is rather the conduction model that seems to describe partly the giant ER effect. Indeed, as we will see later in this paper, the conduction model leads to a linear variation of the force on the electric field and the experimental study of their giant ER fluid show also a quasi-linear law $\tau(E)$.

**The conduction model.** The conduction model appeared in the 90s. Foulc and Atten [4,6] have proposed a theory to which the conductivities of the different phases (liquid and solid) have a major role in the interaction between particles in ER fluids.

The base of the conduction model is the modelling of the force between two spheres in contact (representative of two particles in the ER fluids). Let's consider two spheres of radius r (Fig. 6), immerged in a dielectric liquid and subjected to an electric field E. $\varepsilon_S$, $\varepsilon_L$, $\sigma_S$, $\sigma_L$ are the permittivities and the conductivities of the sphere and the liquid respectively.

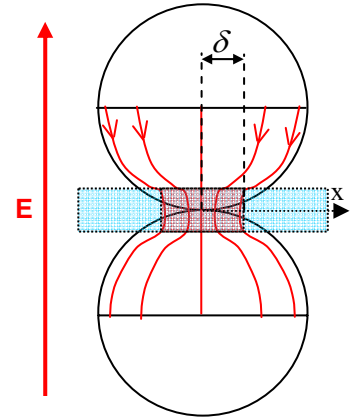

**Figure 6:** Schematic view of two spheres in contact. Sketch of current lines in solid media (x > δ) and in liquid media (electrical contact zone, for x < δ).

The electric field increases drastically as we approach the contact between the spheres. According to Onsager [7], from a certain distance (δ in Fig.6), the conductance of the liquid equal that of the solid and a leakage current cross the liquid between the two sphere surfaces. Two areas can be distinguished. In the first area (x > δ), $\sigma_{solid} > \sigma_{liquid}$ and the current passes through the sphere. In the second area (x < δ), $\sigma_{solid} < \sigma_{liquid}$ and the current goes through the liquid.

Using this hypothesis, the conduction model leads to two laws of the attraction force between spheres as a function of the electric field:

- At low electric fields (<1 kV/mm), the variation of the force with the electric field is quadratic:
$$F \propto E^2 \qquad (1)$$

- For higher electric field (>1kV/mm), the force varies linearly with the field:
$$F \propto E \qquad (2)$$

In both relations (1) and (2) for low and high electric field, a term called $\Gamma = \sigma_s / \sigma_L$ appears, showing the influence of the conductivity of the different phases on the force between the two spheres.

The conduction model has good agreement with experiment as we also find two modes [8] and it works properly for micro sized particles. However, it is not sufficient enough to explain the high yield stress value of giant ER fluids. Other interactions seem to increase the strength of these fluids.

An explanation for the unexpected high value of the yield stress of giant ER fluids could be found in the van der Waals forces. It indeed seems that the giant ER effect appears for nanoscaled molecules and van der Waals forces are rather short-range interaction (in comparison with electrostatic or chemical interactions).

**A new model for the giant ER effect**

**Van der Waals forces.** Even if the van der Waals forces are very weak, they can become powerful as the size of the particles decreases.

We can find in literature [9] an approximation of the van der Waals forces betweens two spheres:

$$F = -\frac{Ar}{12h^2} = -\frac{Ad}{24h^2} \quad (3)$$

where A is the Hamaker constant for the considered media, r the radius (d the diameter) of the particles and h the distance between the two spheres. This force corresponds with the sum of all the interactions between molecules at the surface of the particles (cf. Fig. 7).

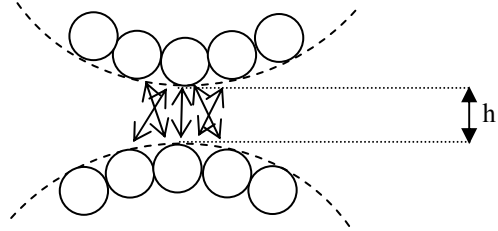

**Figure 7:** van der Waals interaction between two particles

Taking the usual value $10^{-20}$ J for the Hamaker constant, 50 nm as radius and 0.3 nm between the two spheres leads to F ≈ $10^{-9}$ N. This value is the force between two spheres of a single fibre. We can consider that this force corresponds to the force of the whole fibres. Therefore the total force is the sum of the force due to all the fibres.

**Number of fibres.** If we consider our particles as perfect spheres of diameter d, the number of these spheres is given by

$$N_s = \frac{6 \cdot \Phi \cdot Vt}{\pi \cdot d^3}. \quad (4)$$

where $\Phi = V_{solid}/Vt$ is the volume fraction (the ratio of the volume of the particles with the total volume $Vt = V_{solid} + V_{liquid}$).

In the case of two plane electrodes separated from a distance e, $N_{sf} = e/d$ particles are needed to link the two electrodes. Therefore, the number of fibres is

$$N_f = \frac{N_s}{N_{sf}} = \frac{6 \cdot \Phi \cdot Vt}{\pi \cdot d^3} \times \frac{d}{e} = \frac{6 \cdot \Phi \cdot Vt}{\pi \cdot d^2 \cdot e}. \quad (5)$$

and the static yield stress is

$$\tau_s = \frac{F \times Nf}{S} = F \times \frac{6 \cdot \Phi \cdot Vt}{\pi \cdot d^2 \cdot e} \cdot \frac{e}{Vt} = \frac{Ad}{24h^2} \times \frac{6 \cdot \Phi}{\pi \cdot d^2} = \frac{A \cdot \Phi}{4 \cdot \pi \cdot h^2} \times \frac{1}{d} \quad (6)$$

The biggest atom of urea (the surface layer of giant ER particles) is nitrogen whose van der Waals radius is 0.15 nm [10]. We can suppose that the minimum gap between two particles will therefore be h = 0.3nm.
For Φ=0.3, d=100nm and h=0.3 nm we find τ ≈ 26 kPa.
For Φ=0.3, d=50nm and h=0.3 nm we find τ ≈ 53 kPa.

Eq. 6 shows that if the volume fraction and the total volume of fluid are constant, the static yield stress increases when the size of the particles decrease. We can remark that the $\tau_s$ is strongly dependent with the gap between the spheres.

We are far from the 130 kPa but other parameters could improve the theory. This result shows that reason for the so high yield stress of the giant ER fluids can be found in the van der Waals forces. The softness of the particles certainly increases the contact area as the particles crush again each other contributing to a higher van der Waals interaction.

**Conclusion.** This investigation shows the connection between the size of the particles and the yield stress. The van der Waals forces being a plausible cause of the giant ER effect. In this theory, the electric field would be use for bringing the particles closer enough for the van der Waals forces to appear. When the field is turned off, we can suppose that the brutal variation of the force act like a released spring whose energy is high enough to separate the particles.

A more accurate equation of the van der Waals forces have to be investigated, taking into account the smoothness of the particles, a more precise Hamaker constant and the shape as well as the surface of the particles.